\documentclass{article}
\usepackage[latin9]{inputenc}
\usepackage{amsmath}
\usepackage{graphicx}

\makeatletter

\providecommand{\tabularnewline}{\\}

\usepackage{mlspconf}
\copyrightnotice{978-1-5090-6341-3/17/\$31.00 {\copyright}2017 IEEE}

\toappear{2017 IEEE International Workshop on Machine Learning for Signal Processing, Sept.\ 25--28, 2017, Tokyo, Japan}

\title{Inferring Room Semantics Using Acoustic Monitoring}
\name{Muhammad Ahmed Shah, Bhiksha Raj, Khaled A. Harras}
\address{Carnegie Mellon University\\
5000 Forbes Ave, Pittsburgh, PA 15213}
\ninept

\@ifundefined{showcaptionsetup}{}{%
 \PassOptionsToPackage{caption=false}{subfig}}
\usepackage{subfig}
\makeatother

\begin{document}
\maketitle 
\begin{abstract}
Having knowledge of the environmental context of the user i.e. the
knowledge of the users' indoor location and the semantics of their
environment, can facilitate the development of many of location-aware
applications. In this paper, we propose an acoustic monitoring technique
that infers semantic knowledge about an indoor space \emph{over time,}
using audio recordings from it. Our technique uses the impulse response
of these spaces as well as the ambient sounds produced in them in
order to determine a semantic label for them. As we process more recordings,
we update our \emph{confidence} in the assigned label. We evaluate
our technique on a dataset of single-speaker human speech recordings
obtained in different types of rooms at three university buildings.
In our evaluation, the confidence\emph{ }for the true label generally
outstripped the confidence for all other labels and in some cases
converged to 100\% with less than 30 samples.
\end{abstract}

\section{Introduction}

During the last decade, location-based services (LBSs) have become
evermore pervasive with more than 153 million \cite{marketResearch}
users relying on their smartphones for directions, recommendations,
and other location related information. However, most of these LSBs
only function in outdoor environments due to the limitations of the
current localization and mapping technologies. Given that Americans
spend approximately 90\% \cite{klepeis2001national} of their time
indoors and that the market for location-based services is projected
to exceed \$77 billion by 2021 \cite{emarketer_2016}, developing
systems that would facilitate the development of \emph{indoor} LBSs
is a worthwhile endeavor.

A key per-requisite for developing rich indoor LBSs is the ability
to obtain information about the user's environmental context without
any explicit human intervention. The environmental context would include
the user's location within an indoor environment (e.g. second floor,
room opposite the elevator) as well as semantic information about
the surrounding space (e.g. bathroom). This information could then
be used by applications to adapt their behavior, for example, by switching
to silent mode if the user is in an auditorium, or automatically declining
calls if the user is in the bathroom.

The recent developments in the areas of indoor localization \cite{kumar2014ubicarse,levi1996dead,bahl2000radar,mtibaa2015exploiting,abdellatif2013greenloc}
and indoor mapping \cite{alzantot2012crowdinside,gao2014jigsaw,saeed2017argus}
have made it possible to generate accurate indoor floor plans and
track users as they move within them. However, these floor plans are
largely devoid of any semantic labels for the comprising spaces and
therefore have limited usefulness. A few works have attempted to leverage
multiple data sources such as wireless signals, sound, light and sensor
data to associate semantic meanings with indoor locations \cite{bao2015pinplace,azizyan2009surroundsense}
while others have tried to leverage social-network data, such as check-ins,
to annotate unlabeled floor plans with semantic labels \cite{elhamshary2015semsense}.
The infrastructure requirements and training overheads of these approaches
limit their feasibility to be deployed as part of a ubiquitous system.

To offer a more robust, scalable, and practical technique for inferring
the semantics of indoor spaces, we have chosen to rely on audio recordings
obtained from these spaces. While there are several other modalities
that can be employed for this purpose, including wireless signals
and images, certain properties of audio signals make them particularly
convenient to be used in this context. Audio data is easy to collect,
process, and store due to the popularity of handheld devices sporting
high quality recording hardware, and low bandwidth of the audio signal.
Furthermore, audio signals are relatively invariant to changes in
the position and orientation of the recording device, and unlike visual
data, are not limited by ambient lighting conditions. More significantly,
audio signals carry potentially discriminative signatures of the environment
they were recorded in. These characteristics make sound well suited
as a modality to be incorporated in a lightweight semantic inference
system.

In this paper, we propose an acoustic monitoring technique that infers
semantic knowledge about an indoor space by collecting and analyzing
audio recordings from it \emph{over time}. Current approaches for
extracting semantic information present in the acoustic forensics
\cite{zhao2013audio} and scene classification \cite{lee2010detecting,DCASE,malkin2005classifying}
literature predominantly rely on performing \emph{one-off} classification
in which each individual data sample is classified independently.
Meanwhile, our approach to this problem is fundamentally different,
and to the best of our knowledge, has not been explored in the current
literature. We treat individual audio samples as evidence to help
us infer a semantic label for the environment it was obtained from.
Like in a forensic investigation, a single piece of evidence is of
limited significance. As part of a robust investigative process, all
the evidence must be aggregated for inferences to be drawn from it.
As we obtain and analyze more samples we accumulate more evidence
and our confidence in the label we inferred for the space increases.
Summarizing the novelty of our approach, when analyzing an audio recording,
the current approaches try to answer the question ``where was this
recorded?'', whereas we ask a different question altogether, i.e.
``given this recording, and $n$ recordings that we had received
earlier from the \emph{same place}, how likely is it that this place
is, for example, an office?''.

For inferring the semantics of a space, our technique uses appropriate
models to extract and combine evidence from two acoustic characteristics
of the space, namely, ambient sounds and the impulse response. We
use Mel-Frequency Cepstral Coefficients (MFCCs) to model the ambient
soundscape. While, on the other hand, we use a non-negative de-convolution
approach to isolate the Room Impulse Response (RIR). We then use Gaussian
Mixture Models (GMM) and Support Vector Machines (SVMs) to classify
the MFCCs and RIR respectively . We find that both of these characteristics
are capable of independently providing confirmation of the semantics
of a space. Moreover, we also find that combining evidence from both
yields significantly better performance than either of the characteristics
alone. As mentioned above, our approach incrementally builds up confidence
in the classification of a room over time from the accumulated evidence
of multiple recordings. It does so by aggregating the evidence provided
by our models using Bayesian inference. 

We evaluated our approach on a large dataset of audio recordings.
We collected over 12,000 recordings from five different types of room
in three university buildings using smartphones. Our technique was
able to correctly identify all the room types when the training and
testing data came from the same set of buildings. When tested on an
unseen building it only misclassified one room type. In our evaluations,
our technique was able to reach 100\% classification confidence with
as less as 30 recordings.

The remainder of the paper is organized as follows, section 2 presents
the feature extraction techniques and learning models we have used,
section 3 presents the evaluation process and the results and the
discussion in section 4 ends the paper.

\section{Inferring Room Semantics from Audio}

Our objective is to automatically detect room semantics \textendash{}
the purpose of any room \textendash{} from audio recordings within
that room. Audio signals have a convenient property of being able
to capture potentially discriminatory information about the environment.
From a purely acoustic perspective, environments, rooms in our case,
can differ along two dimensions: (1) The acoustic content, that characterizes
the typical sounds in the room and (2) the room impulse response,
which characterizes its structure. 

Our approach employs different pre-processing and classification strategies
to exploit each of these characteristics and combine them for our
final decision. We extract MFCCs from the audio recordings then use
GMMs to model the ambient soundscape present in each room type. We
also extract the RIR from the Short-Time Fourier Transform (STFT)
using the non-negative deconvolution approach below. We parameterize
the RIR and use it to train SVM classifiers for each room type. Finally,
we must also consider that a room may serve multiple purposes. For
instance, a pantry may double as a printer room in an office building.
Thus, rather than attempting to {\em classify} between multiple
semantic options, we will treat the problem as one of {\em detection},
applied to each semantic. The decision is not intended to be instantaneous;
rather, we will attempt to determine the confidence with which each
semantic may be assigned to the room incrementally, from recordings
obtained over time. As a result, we require a higher-level framework
within which evidence is combined to incrementally build up confidence
in the label attributed to the room.

We describe each of the components of our procedure below. 

\vspace{-10bp}

\subsection{Classifying by Acoustic Content}

Ambient sounds can be used to qualitatively differentiate a room.
These ambient sounds can be thought of as composing an acoustic scene
that is intrinsically linked to some high-level semantic of the space.
For example, the sound of water splashing is a characteristic of the
bathroom, while a kitchen space may be expected to contain the sounds
of appliances humming.

We cast the problem of identifying a room type as one of {\em acoustic
scene classification} \cite{barchiesi2015acoustic}. In modeling
the ambient sound scene, the temporal arrangement of sound patterns
is not of particular significance, rather what we are interested in
is capturing the qualitative features of the sound scene. While a
number of techniques have been proposed for this purpose in the literature
\cite{barchiesi2015acoustic,geiger2013large,giannoulis2013detection},
in our work we have chosen to use a Gaussian mixture classifier. We
do so because it has been shown that the qualitative features of an
acoustic scene can be effectively modeled by the distribution of frame-based
spectral features \cite{aucouturier2007bag}. Furthermore, this approach
also enables us to compute log-likelihoods, which is convenient for
subsequent calculations. 


For each room type $C$ we model the distribution of Mel-Frequency
Cepstral vectors derived from acoustic recordings in instances of
that room as a Gaussian mixture: 
\begin{equation}
p(\vec{x}|C)=\sum_{i=1}^{N}w_{C,i}{\mathcal{N}}(\vec{x};\mu_{C,i},\Sigma_{C,i})\label{eq:gm}
\end{equation}
, where $\vec{x}$ represents a random cepstral vector, ${\mathcal{N}}(\vec{x};\mu,\Sigma)$
represents a Gaussian distribution for $\vec{x}$ with mean $\mu$
and variance $\Sigma$, $N$ represents the number of Gaussians in
the mixture, and $w_{C,1},\cdots,w_{C,N}$ are the weights of the
Gaussians in the mixture. Similarly, the probability density function
of vectors from recordings that do {\em not} belong to the room
are also modeled by a Gaussian mixture obtained by replacing $C$
by $\neg C$ in equation (\ref{eq:gm}) where $\neg C$ is the set
of all rooms types except $C$. The parameters of both distributions
(namely the mixture weights, means, and variances) are learned from
training instances of recordings using the expectation-maximization
algorithm \cite{dempster1977maximum}. 

Given a test recording $\mathbf{X}=\vec{x}_{1},\cdots,\vec{x}_{T}$,
we compute the {\em within class} and {\em out-of-class} log
likelihoods as $L_{A}(\mathbf{X}|C)=\sum_{t}\log p(\vec{x}_{t}|C)$,
and $L_{A}(\mathbf{X}|\neg C)=\Sigma_{t}\log p(\vec{x}_{t}|\neg C)$.
Classification may be performed by directly comparing $L_{A}(\mathbf{X}|C)-L_{A}(\mathbf{X}|\neg C)$
to a threshold. We will, however, use these values to compute confidences
as explained below.

\subsection{Room Impulse Response Extraction and Classification}

The purpose of a room is also reflected in the size, shape, and furnishings
of the room. These, in turn, will affect the {\em Room Impulse Response}
of the room. For instance, the reverberation times and the room response
of bathrooms, which are generally small-to-mid sized, devoid of furniture,
and often have hard reflective surfaces such as tiles and mirrors,
will be very different from those of a class room with tables, chairs,
and (usually) people occupying them, and these in turn will be different
from offices with their very different form factors and furnishings.

In order to exploit this information to classify rooms, however, it
will be necessary to {\em derive} the room response from the recordings,
parametrize them appropriately, and utilize an appropriate classifier.
We discuss our approach below.

\subsubsection{Extracting Room Impulse Response}

\label{de-reverb}

Extracting exact room responses from a monaural recording remains
a challenging and unsolved problem. However, for our purposes, an
approximate estimate obtained from general principles will suffice.
We use the non-negative deconvolution method described in \cite{kameoka2009robust}
for our purpose.

Following the approach of \cite{kameoka2009robust}, we model the
effect of the room on the acoustic signal as a convolution of the
magnitude spectrogram of the signal with the magnitude spectrogram
of the room impulse response. The room impulse response is then extracted
from the magnitude spectrogram of the reverberant recorded signal
through non-negative matrix deconvolution.

Given an observed signal, represented by a $F\times T$ magnitude-spectrogram
matrix, $\mathbf{X}$, (where $T$ represents the number of time frames
and $F$ is the number of frequency bands), our goal is to approximate
$F\times T$ matrix, $\mathbf{S}$ and $F\times K$ matrix $\mathbf{R}$,
such that $\mathbf{Y=S*R}$ and $\mathbf{Y\simeq X}$, $*$ represents
the convolution operation and, $\mathbf{S}$ and $\mathbf{R}$ represent
the magnitude spectrograms of the speech and RIR respectively.

We define $F\times F$ matrices $\mathbf{R}_{1},\ldots,\mathbf{R}_{K}$
such that 
\[
\mathbf{R}_{k}=diag(R_{1}[k],...,R_{F}[k])
\]
where $\mathbf{R}_{k}$ is a diagonal matrix with $R_{1}[k],...,R_{F}[k]$
as the values on the diagonal and $R_{f}[t]$ represents the magnitude
spectral component of the $f^{th}$ frequency sub-band at time index
$t$ for the RIR. Now we can formulate the approximation as 
\[
\mathbf{Y=\sum_{\text{\ensuremath{k=1}}}^{\text{\ensuremath{K}}}R_{\text{\ensuremath{k}}}\cdot\overset{\text{\ensuremath{k\rightarrow}}}{S}}.
\]
where $k\rightarrow$ represents a right shift of the spectrogram
by $k$ time steps. Since we are dealing with speech, $\mathbf{S}$
needs to be sparse. To encourage sparsity we use the update rule for
$\mathbf{S}$ presented in \cite{kameoka2009robust}, which introduces
sparsity parameters $p$ and $\lambda$. The update rules for $\mathbf{S}$
and $\mathbf{R}$ are
\[
\mathbf{S}\leftarrow\mathbf{S}\otimes\dfrac{\mathbf{\sum_{\text{\ensuremath{k=1}}}^{\text{\ensuremath{K}}}R_{\text{\ensuremath{k}}}\cdot\overset{\text{\ensuremath{k\leftarrow}}}{X}}}{\mathbf{\lambda}|\mathbf{S}|^{p-1}+\mathbf{\sum_{\text{\ensuremath{k=1}}}^{\text{\ensuremath{K}}}R_{\text{\ensuremath{k}}}\cdot\overset{\text{\ensuremath{k\leftarrow}}}{Y}}}
\]
\[
\mathbf{R_{\text{\ensuremath{k}}}=R_{\text{\ensuremath{k}}}\otimes\dfrac{\dfrac{X}{Y}\cdot\overset{\text{\ensuremath{k\rightarrow}}}{S}^{\top}}{1\cdot\overset{\text{\ensuremath{k\rightarrow}}}{S}^{\top}}},
\]
where $\otimes$ represents the Hadamard product. After the procedure
has converged we reconstruct a $F\times K$ matrix, $\mathbf{R}$
representing the RIR by placing the diagonal of $\mathbf{R}_{k}$
at the $k^{\text{th }}$ column of $\mathbf{R}$.

\subsubsection{Parametrizing and Classifying the RIR}

The outcome of the above procedure is a $F\times K$ matrix representing
the magnitude spectrogram of the RIR. Note that although the actual
RIR of any room is infinite in length, our estimates returns a finite
and fixed-length RIR; this approximation enables us to obtain fixed-
and finite-length characterizations that we can use for classification.

We compress the RIR further by taking the logarithm of the magnitude
spectra and performing a Discrete Cosine Transform (DCT) on individual
rows. We only keep the first 20 cepstral features from the DCT, so
that we are left with a $20\times K$ matrix for the RIR. Since the
temporal pattern represented by the arrangement of the rows is significant
we simply flatten this matrix to obtain a $20K$-dimensional feature
vector. Note that this way we retain the {\em entire} RIR as a
feature and in doing so we account for the temporal structure of the
audio frames in the RIR. 

To classify the RIR, we use Support Vector Machines (SVMs). We chose
SVMs, instead of using distribution models such as i-vectors or GMMs,
for this task because, unlike the acoustic scene, classifying RIR
requires us to explicitly account for the temporal variations in the
signal, making distribution-based method unsuitable. We setup the
SVM to map the classification scores to values between 0 and 1 and
can be interpreted as probabilities. Let $L_{R}(\mathbf{X}|C)$ and
$L_{R}(\mathbf{X}|\neg C)$ represent the {\em log} probabilities
assigned to $C$ and $\neg C$ by the SVM. In order to classify the
RIR, we can compare the difference of the two log probabilities to
a threshold; however, as in the case of scene classification described
above, we will use them to compute confidences.

\subsection{Incremental Confidence Calculation}

\label{confidence} Finally, we compose the results from the individual
recordings to obtain an incrementally updated confidence value for
each room type $C$. In principle, these could be computed directly
through iterated computation of {\em a posteriori} probabilities
of the classes via Bayes rule using the original probability values
obtained from the GMMs and SVM. Instead, we employ the approach described
in \cite{singh2004classification} as we find it to be more effective.

First we combine the results from the GMMs and SVMs into a \emph{hybrid}
model. Each model is designed to capture a specific set of features
from the recordings so relying exclusively on one of them may lead
to potentially relevant information being discarded. While on the
other hand by combining the two classifiers we stand to benefit from
the unique information captured by each model. For each recording,
$\mathbf{X}$, we compute log-likelihood ratio scores, $\mathbf{L}_{A}=L_{A}(\mathbf{X}|C)-L_{A}(\mathbf{X}|\neg C)$
and $\mathbf{L}_{R}=L_{R}(\mathbf{X}|C)-L_{R}(\mathbf{X}|\neg C)$,
from the acoustic scene and RIR classifiers respectively. We then
compute a single overall log-likelihood ratio score given by 
\begin{equation}
p_{i}=\alpha(\mathbf{L}_{A})+(1-\alpha)(\mathbf{L}_{R})\label{eq:}
\end{equation}
where $\alpha$ weighs the contribution of the acoustic scene classifier
in the overall score. Let $p_{i}'=p_{i}-t_{C}$ where $t_{C}$ is
a constant threshold value. Define $P(p_{i}'|C)$ and $P(p_{i}'|\neg C)$
to be the distributions of the $p_{i}'$ values for $C$ and $\neg C$.
In practice we set $t_{C}$ to be the threshold value at which the
False Positive Rate (FPR) equals the False Negative Rate (FNR) for
class $C$. This value of FPR and FNR is known as the Equal Error
Rate (EER). Following \cite{singh2004classification} we model both
as Gaussian Mixtures, the parameters of which can be trained from
training data. Given a sequence of observations $p_{1},\cdots,p_{n}$
and the respective $p_{i}'$, the incremental confidence update rule
is given by 
\begin{equation}
P(C|p_{1}',\cdots,p_{n}')=\frac{\Pi_{i}^{n}P(p_{i}'|C)}{\Pi_{i}^{n}P(p_{i}'|C)+\omega\Pi_{i}^{n}P(p_{i}'|\neg C)}\label{eq:-1}
\end{equation}
where $\omega$ is a constant that represents $\dfrac{P(\neg C)}{P(C)}$.

\section{Evaluation}

\subsection{Dataset}

We have compiled a data set comprising over 12,004 speech recordings
for our experiments. The speech is recorded by two smartphones simultaneously
over a single channel at a sampling rate of 44.1 Khz. A Samsung Galaxy
SIII mini is held by the speaker at chest level, while a Sony Xperia
Z1 Compact is in his pocket. The recordings were performed in several\emph{
Bathrooms}(BR), \emph{Offices}(O), \emph{Pantries}(P)\emph{, Classrooms}(C)
and \emph{Lecture Halls} (LH) on three university buildings (referred
to as C-I, C-II and C-III below). In each room we took 100 recordings,
3 seconds each. The speaker stood at 5 locations within the room and
uttered 4 short sentences, repeating each sentence 5 times. The recordings
were obtained at times when there was low foot traffic in order to
keep them as clean as possible.

For the experiments in this paper, we use only the recordings from
the aforementioned dataset in which the smartphone was held up by
the speaker, since the recordings obtained with the phone kept in
the speaker's pocket could not have effectively captured the RIR.

We divide the data into four folds for cross-validation. All experimental
results reported below have been averaged over the 4 folds. The folds
are constructed such that no two folds have data from the same room
in order to ensure that the training set does not have any data from
the rooms in the test set.

\subsection{Feature Extraction}

We extract two feature sets from the raw audio signal, Mel-Frequency
Cepstral Coefficients (MFCCs) and the RIR. For both, we compute the
Short Time Fourier Transform (STFT) using a 64ms hamming window with
32ms overlap. To model ambient sound we compute 20-dimensional MFCCs,
extended with difference coefficients \cite{reynolds1995robust}.
To extract the parameterized RIR we apply the NMFD based de-reverberation
procedure described earlier to STFT.

\subsection{Training}

For scene classification we train two Gaussian mixtures, each having
64 components, per room type. One mixture is trained on the data for
the specific room type while the other is trained on the data from
all other room types. To classify the RIRs, we trained a binary SVM
with a radial-basis function (RBF) kernel for each room type using
LIBSVM\cite{chang2011libsvm}. We setup the SVM to output probability
values for the positive and negative classes and used grid search
with five fold cross-validation to find the optimal parameters for
the SVM.

\vspace{-2.5bp}

\subsection{Results}

\subsubsection{Classifier Evaluation }

\begin{table}
\centering %
\begin{tabular}{|c|c|c|}
\hline 
{Room Type}  & {\emph{Scene}}  & {\emph{RIR}} \tabularnewline
\hline 
{}Lecture Hall  & 8.72 & 13.53\tabularnewline
\hline 
{}Classroom  & 12.23 & 34.03\tabularnewline
\hline 
{}Pantry  & 10.68 & 22.05\tabularnewline
\hline 
{}Office  & 6.42 & 23.23\tabularnewline
\hline 
{}Bathroom  & 8.2 & 5.5\tabularnewline
\hline 
\end{tabular}

\caption{\label{tab:Equal-Error-Rate}EER, acoustic scene classification.}
\end{table}

We first evaluate room classification performance on {\em individual}
recordings obtained from campuses C-I and C-II, using the equal error
rate (EER) as the metric. It is important to note that while the testing
and training set do not have data from the same rooms, they do have
data from the same set of buildings. The EERs for the acoustic scene
and RIR classifiers are given in Table \ref{tab:Equal-Error-Rate}.
While both classifiers produced encouraging results, the acoustic
scene classifier outperformed the RIR classifier in all room types
except for the bathrooms suggesting that the acoustic scene is well
suited for characterizing rooms of different types. Though less impressive
than the acoustic scene classifier, the results from the RIR classifier
are still quite good. The RIR classifier performs very well on environments
with distinctive structural features, namely lecture halls and bathrooms.
The high ceilings of lecture halls and the ceramic tiling in the bathrooms
would produce acoustic artifacts which are not found in other types
of rooms. However, we see that structural information alone, as obtained
from the RIR, may not be sufficient when trying to classify more complex
acoustic environments such as classrooms, offices and pantries. With
that said, we still maintain that the structural information captured
by the RIR is salient in the semantic inference process. As we shall
see in the next section, augmenting the acoustic scene classifier
with the structural information from the RIR yield significantly better
classification performance.

\begin{figure}
\begin{centering}
\includegraphics[scale=0.45]{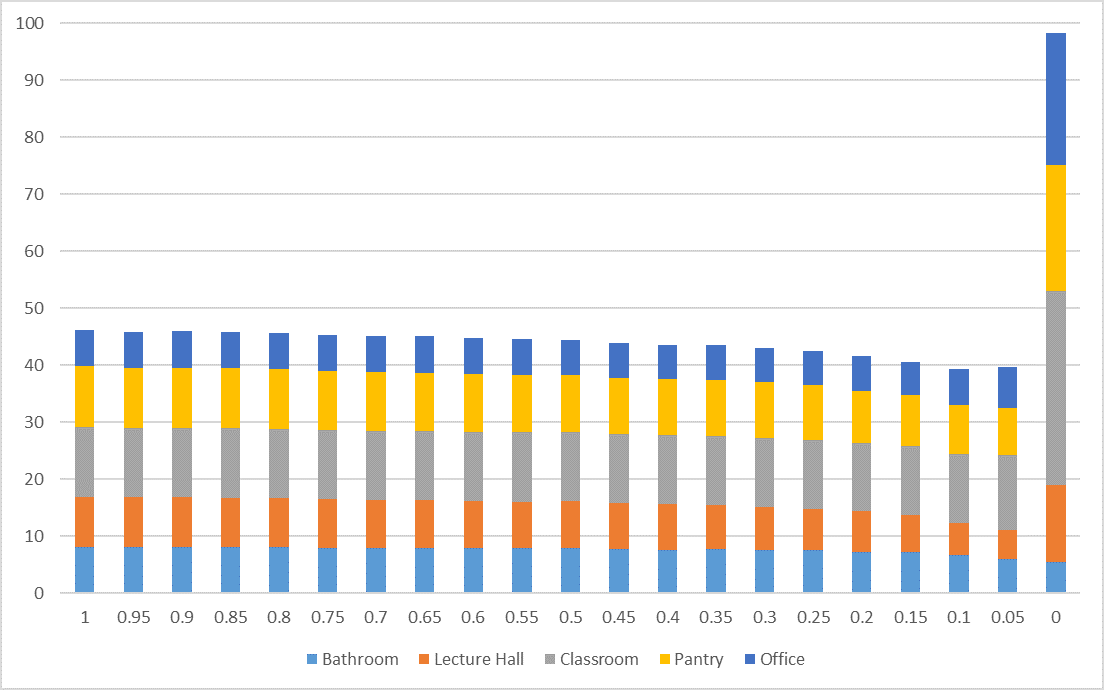}
\par\end{centering}
\caption{{\small{}\label{fig:Changes-in-EER}Total EER as $\alpha$ is varied.
Each color represents one type of room.}}
\end{figure}

\subsubsection{Combining The Classifiers}

As we stated in \ref{confidence}, we want to aggregate the information
obtained from both, the acoustic scene and the RIR classifiers into
a \emph{hybrid} model using equation (\ref{eq:}). We empirically
determine the optimal value of the weight parameter $\alpha$ in equation
(\ref{eq:}) by computing the total EER, across all room types, for
different values of $\alpha$ and choosing the $\alpha$ at which
the total EER is minimized.

Fig.\ref{fig:Changes-in-EER} illustrates how the total EER decreases
as $\alpha$ is varied. The EER decreases as $\alpha$ is decreased
reaching a minimum at $\alpha=0.10$. In fact, the total EER obtained
by combining the classification results, with $\alpha=0.10$, is 14.7\%
lower than that of the audio scene classifier ($\alpha=1.0$) and
59.9\% lower than that of the RIR classifier ($\alpha=0.0$). This
observation is quite interesting because the RIR classifier alone
yields very high error rates but by providing just a little information
from the acoustic scene classifier we can drastically reduce them
and obtain better performance than the acoustic scene classifier itself.

\subsubsection{Confidence Calculation}

\begin{figure}
\subfloat[Lecture Halls]{\includegraphics[scale=0.22]{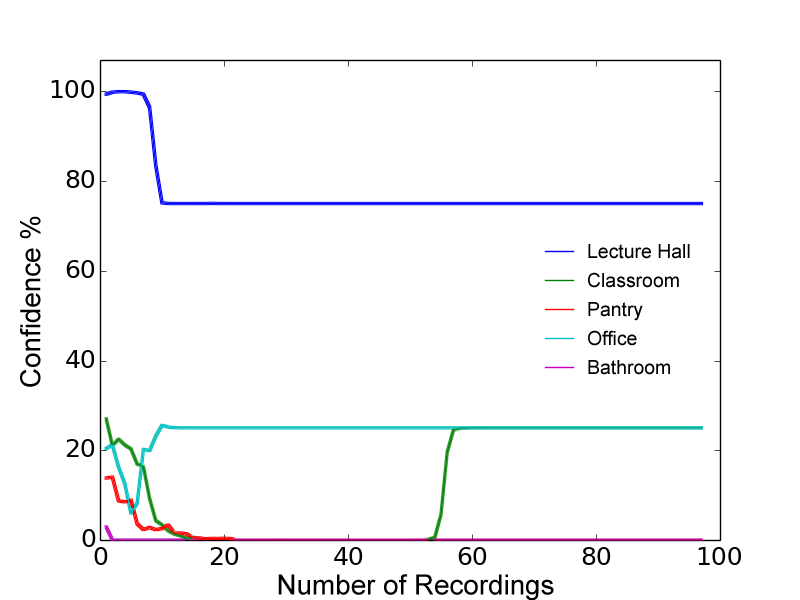}

}\subfloat[Offices]{\includegraphics[scale=0.22]{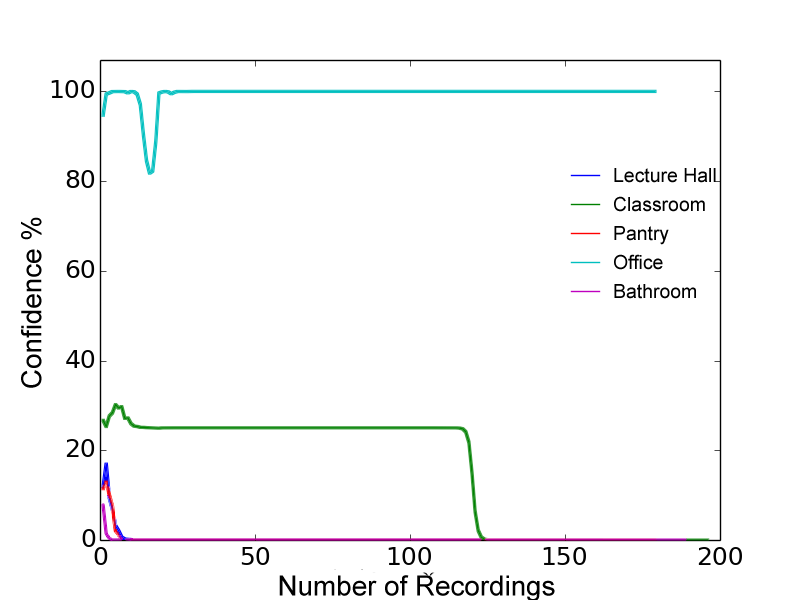}

}

\subfloat[Pantries]{\includegraphics[scale=0.22]{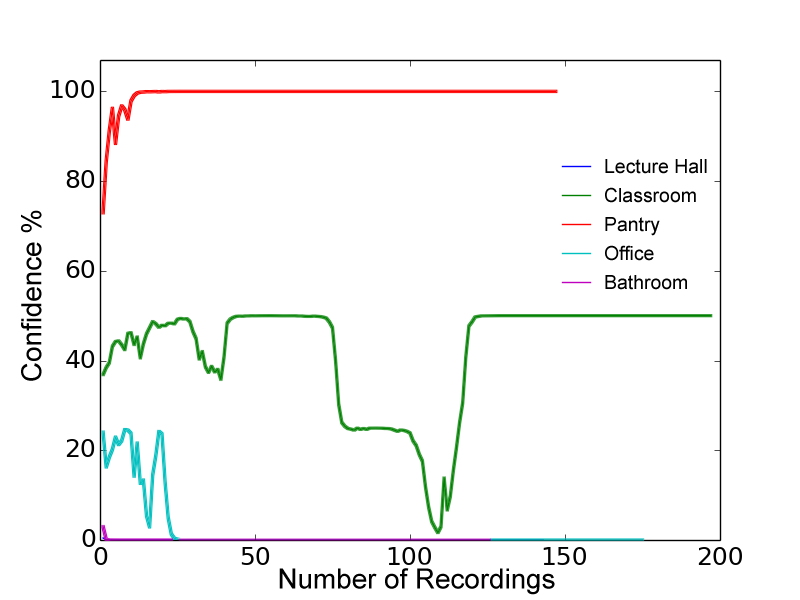}

}\subfloat[\label{fig:bathroom} Bathrooms]{\includegraphics[scale=0.22]{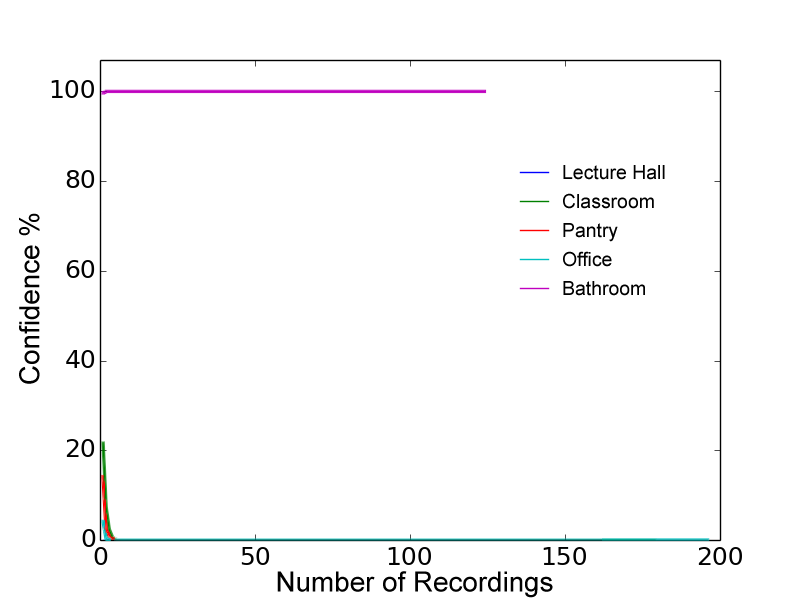} 

}

\caption{{\small{}\label{fig:Confidence-buildup-as}Confidence build up per
room using the hybrid classifier with $\alpha=0.10$}}

\vspace{-10bp}
\end{figure}

We evaluate the ability of the proposed approach to incrementally
buildup confidence in semantic labels for the rooms. Given a set of
recordings, all obtained from within the same room, we use the Bayesian
inference procedure described in \ref{confidence} to incrementally
update our confidence in the candidate semantic labels for the given
room. As more audio samples are made available to the classifier we
expect to see the confidence in the true label increase while the
confidence in other labels diminishes. In our experiments we set $\omega=4$
and, model $P(p_{i}'|C)$ and $P(p_{i}'|\neg C)$ as a four component
Gaussian mixture. Figure \ref{fig:Confidence-buildup-as} illustrates
the process of confidence build up on the data from campuses C-I and
C-II when classified using the \emph{hybrid} model with $\alpha=0.10$.

It is encouraging to see that our technique can confidently infer
the correct semantic label for different room types given a set of
audio recordings. For all the room types in our dataset, our confidence
in the true label clearly outstripped our confidence in all other
labels. For three out of five room types the confidence in the true
label rapidly approached 100\%, with \emph{less than 30 samples} and
thereby demonstrating the practical viability of our proposed approach.
By far the most impressive performance was shown by bathrooms which
were unambiguously labeled with 100\% confidence with less than a
handful of recordings. The performance was not as impressive for lecture
halls but confidence in the true label still converged to more than
75\% with less than 20 samples, while the confidence for all other
labels remained much lower. 

Another interesting consequence of our technique is that the confidence
values for different rooms are not correlated. This permits our technique
to accurately represent situations in which a room serves multiple
purposes and needs to be assigned more than one semantic tag or, if
the room is of a type that is not known to our classifier, to be flagged
as unknown.

\subsubsection{Testing On An Unseen Building}

While the results in Fig. \ref{fig:Confidence-buildup-as} are very
encouraging they are not conclusive because the rooms in the training
and testing set come from the same set of building. It is very common
for rooms of a particular type in the same buildings to be homogeneous
in their structure and layout. This homogeneity can lead to similar
acoustic features being observed in these rooms. Therefore, we run
the risk of our models over-fitting to these common features and producing
overly optimistic results. 

\begin{figure}
\begin{centering}
\includegraphics[scale=0.25]{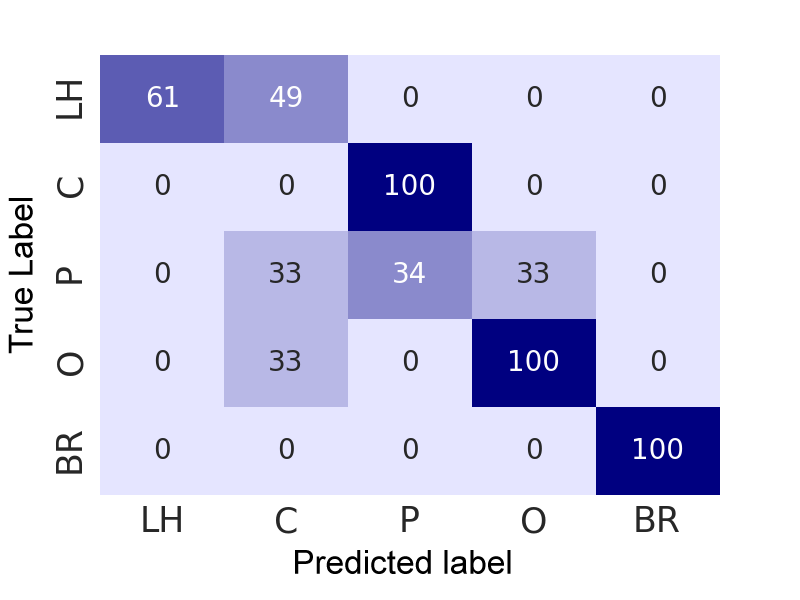}
\par\end{centering}
\caption{{\small{}\label{fig:confMat}Confusion Matrix of final classification
confidence values on the unseen building.}}

\vspace{-10bp}
\end{figure}

To obtain a more reliable set of results we evaluate our technique
on an ``unseen'' building i.e. a building not included in the training
set. We train our classifiers using the data from the campuses C-I
and C-II and test on the data from the campus C-III. We use the \emph{hybrid}
classification model with $\alpha=0.10$ and calculate confidence
using equation (\ref{eq:-1}) with thresholds, $t_{C}$, set to threshold
values corresponding to the EER values in Table \ref{tab:Equal-Error-Rate}.
Due to space considerations we have chosen to only illustrate the
final confidence values for each label in the form of a confusion
matrix presented in Fig.\ref{fig:confMat}.

Notwithstanding the fact that our training dataset only had data from
two buildings and hence was lacking in diversity, our technique was
able to correctly infer the labels for four out of the five room types
in our dataset. We were able to confidently label lecture halls, bathrooms
and offices. While we were absolutely certain about our labels for
the offices and bathrooms, the lecture halls were being confused with
classrooms in some cases. This is not a serious setback since lecture
halls and classrooms can have a lot of similar features and their
structural layout can vary significantly across different buildings.
However, our technique was not very sure about the label for pantries
and grossly miss-classified the classrooms. We believe that these
errors were caused by the fact that the classrooms and pantries in
the test building significantly differed in their layout and content
from their counterparts in the training buildings. Unlike the training
classrooms, the classrooms in the test buildings also served as computer
clusters while the pantries in the test building were significantly
smaller than the ones in the training set and lacked certain appliances
such as printers. These differences would probably have changed the
ambient sound patterns present in the rooms as well the impulse response.

\section{Conclusion}

We have proposed an acoustic monitoring scheme to detect room semantics
from audio recordings. We have demonstrated that both qualitative
(ambient sound scene) and structural (RIR) identifiers of spaces are
capable of providing confirmation of a room's semantic label, however,
a linear combination of the two classifiers yields lower error rates
than each classifier achieved individually. We have also developed
a Bayesian inference technique for aggregating the evidence obtained
from the classifiers to build up confidence in semantic labels for
rooms over time. Finally we have evaluated our system using audio
recorded in three university buildings. On the validation dataset,
consisting of data from two campuses, the confidence of our system
in the true label significantly outstripped its confidence in all
other labels. Moreover, our system also performed very well on the
data from the third campus, that was not included in a training set,
assigning correct labels to 4 out of 5 different classes of rooms.

The work presented in this paper is part of a larger project to create
a lightweight crowdsourced system for automated annotation of indoor
floorplans, in which audio is one of the modalities, along with visual
and sensory data, to be employed to infer the semantic labels for
rooms. As future work, we plan on performing evaluation trials with
volunteers. We intend to make all the data we gather available to
the public to facilitate future work in this area. 

{ \bibliographystyle{IEEEbib}
\bibliography{mybib}
}
\end{document}